\documentclass[aps,prl,twocolumn,noshowpacs,superscriptaddress,longbibliography]{revtex4}
\usepackage{graphicx}
\usepackage{amsmath,amssymb,amsfonts}

\begin{document}

\title{Demonstration of Hopf-link semimetal bands with superconducting
circuits}
\author{Xinsheng Tan}
\author{Mengmeng Li}
\author{Danyu Li}
\author{Kunzhe Dai}
\author{Haifeng Yu}
\email{hfyu@nju.edu.cn}
\author{Yang Yu}
\email{yuyang@nju.edu.cn}
\affiliation{National Laboratory of Solid State Microstructures, School of Physics,
Nanjing University, Nanjing 210093, China}

\begin{abstract}
Hopf-link semimetals exhibit exotic gapless band structures with fascinating
topological properties, which have never  been observed in nature. Here we
demonstrate nodal lines with topological form of Hopf-link chains in
artificial semimetal-bands. Driving superconducting quantum circuits with
elaborately designed microwave fields, we mapped the momentum space of a
lattice to a parameter space of the Hamiltonian for a Hopf-link semimetal.
By measuring the energy spectrum, we directly imaged nodal lines in cubic
lattices. By tuning the driving fields, we adjusted various parameters of
Hamiltonian. Important topological features, such as link-unlink topological
transitions and the robustness of the Hopf-link chain structure were
investigated. Moreover, we extracted the linking number by detecting the
Berry phase associated with different loops encircling nodal lines. This
topological invariant clearly reveals the nontrivial topology of the
Hopf-link semimetal. Our results provide knowledge for developing new
materials and quantum devices.
\end{abstract}

\maketitle
\affiliation{National Laboratory of Solid State Microstructures, School of Physics,
Nanjing University, Nanjing 210093, China}
\affiliation{Department of Physics and Center of Theoretical and Computational Physics,
The University of Hong Kong, Pokfulam Road, Hong Kong, China}
\affiliation{National Laboratory of Solid State Microstructures, School of Physics,
Nanjing University, Nanjing 210093, China}
\affiliation{National Laboratory of Solid State Microstructures, School of Physics,
Nanjing University, Nanjing 210093, China}
\affiliation{Institute of Physics, Chinese Academy of Sciences, Beijing, China}
\affiliation{National Laboratory of Solid State Microstructures, School of Physics,
Nanjing University, Nanjing 210093, China}
\affiliation{Department of Physics and Center of Theoretical and Computational Physics,
The University of Hong Kong, Pokfulam Road, Hong Kong, China}
\affiliation{National Laboratory of Solid State Microstructures, School of Physics,
Nanjing University, Nanjing 210093, China}




\section*{Introduction}

\begin{figure}[tbph]
\includegraphics[width=7.5cm]{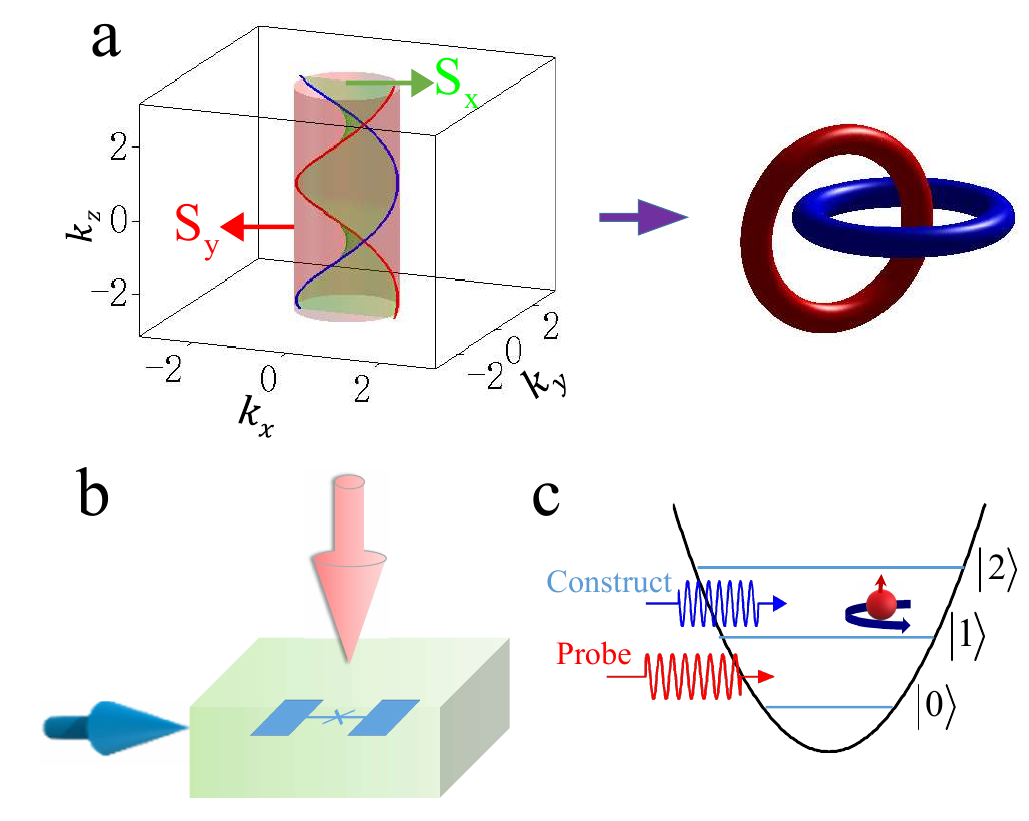}
\caption{(Color online) \textbf{a}, Nodal lines with a double-helix
structure formed by the intersection of two surface $S_1$ and $S_2$. It is
topologically equivalent to a Hopf-link. \textbf{b}, A superconducting
transmon embedded in three-dimensional cavity are driven by designed
microwaves, realizing the effective Hamiltonian to simulate Hopf-link
semimetals. \textbf{c}, The schematic energy structure of a transmon. The
lowest three energy levels are used to do the simulation.}
\label{demon}
\end{figure}

Topology plays a very important role in physics research, inspiring many
findings in condensed matter physics during the past decades. For instance,
topological insulators and superconductors \cite{Hasan_RMP,Qi_RMP}, which
have gapped bulks with topological structures, have been both theoretically
and experimentally discovered recently \cite{XGWan-PRB11,Konig_Hall,Kane_Z2}%
. Their physical properties are characterized by nontrivial topological
invariants. In addition to protected gapped systems, gapless band structures
can also be topological materials, such as Weyl semimetals \cite%
{Lu-Science15,Xu-Science15,LiuDirac,LvWeyl,bian_nodal}. The study of
band-touching manifolds, including semimetals with 0D nodal points (or Dirac
points) and 1D nodal lines, deepens our understanding of condensed matter
physics \cite{
OskarWeyl,YuDiracline,KimDiracline,burkov_nodal,moore_hopf,Neupane-NC15}.
Comparing to 0D gapless modes, nodal lines provide richer topological
structures: nodal rings can touch at special points, resulting in various
shapes of nodal chains. Recently, it was theoretically predicted that
semimetals with a unique Hopf-link structure are possible \cite%
{chen_hopf,yan_nodal-link}. The Hopf-link, which consists of two rings that
pass through the center of each other, represents the simplest topologically
nontrivial link. A typical Hamiltonian based on a cubic lattice with
Hopf-link structure \cite{chen_hopf} is given by
\begin{equation}
\begin{array}{ll}
H(\mathbf{k})=f_1 (\mathbf{k})\sigma _{1}+f_2 (\mathbf{k})\sigma _{3}, &  \\
f_1(\mathbf{k})=\sin k_y \cos k_z - \sin k_x \sin k_z, &  \\
f_2(\mathbf{k})= 2 \cos k_x + 2 \cos k_y +\chi, &
\end{array}
\label{Lattice-model}
\end{equation}
where $\sigma_{1,2,3}$ are the Pauli matrices and $\chi$ is a tunable
parameter. Nodal lines in the Brillouin zone can be interpreted as the
intersecting lines of two surfaces $S_x$: $f_1(\mathbf{k})$ = 0 and $S_y$: $%
f_2(\mathbf{k})$ = 0, forming a novel double-helix structure. Furthermore,
due to the periodicity of the Brillouin zone, the cylinder $S_y$ folds into
a torus, so the double-helix structure deforms into a Hopf-chain [as shown
in Fig.~\ref{demon}(a)]. The nodal loop cannot shrink to a point without
crossing each other, leading to a finite linking number of the Hopf-chain.
This basic topological invariance can be extracted from the Berry phase \cite%
{berry_berry} carried by a closed loop which encircles the nodal rings \cite%
{chang_weyl-link,chen_hopf}. It is easy to verify that Hamiltonians with
Hopf-chain structure obey $PT$ combined symmetry, while breaking the
individual $T$ and $P$ symmetry since $[H(\mathbf{k}),PT]=0$, where $T$ is
the time-reversal operator and $P$ is the spatial-inversion operator.

In this paper, we have experimentally realized the topological Hopf-link
semimetal bands in a square-lattice, via an analogy between the momentum
space with a controllable parameter space in superconducting quantum
circuits. By measuring the whole energy spectrum of our system in Brillouin
zone, we have clearly imaged the gapless band structure of topological
semimetals with linked nodal lines. The topological link-unlink phase
transitions in semimetal bands can be manipulated by intentionally adding an
extra term in the simulated effective Hamiltonian \cite{chen_hopf}.
Furthermore, to demonstrate the topological robustness of the Hopf-link, a
perturbation which breaks $P$ and $T$ symmetry, while preserves joint $PT$
symmetry, is applied. It is verified experimentally that the Hopf-chain of
the topological semimetal bands are still present under such perturbations,
although the position of nodal lines are changed drastically. To
characterize the topological properties of the Hopf-chain, we extracted the
linking numbers of nodal lines by detecting the Berry phase after evolving
the system along designed paths enclosing or disclosing nodal lines
respectively \cite{yan_nodal-link,chen_hopf}. All of these observations
illustrate convincingly the topological properties of Hopf-link semimetals,
making these systems promising to develop new materials and devices.

\section*{Results}

The superconducting quantum circuits used in our experiment consist of a
superconducting transmon qubit embedded in a 3D aluminium cavity~\cite%
{Tan_pt,paik_3d,devoret_3d,shankar_3d,wang_3d}. The transmon qubit, composed
of a single Josephson junction and two pads (250 $\mu $m $\times $ 500 $\mu $%
m), is patterned using standard e-beam lithography, followed by double-angle
evaporation of aluminium on a 500 $\mu $m thick silicon substrate. The
thicknesses of the Al film are 30nm and 80nm, respectively. The chip is
diced into 3 mm $\times $ 6.8 mm size to fit into the 3D rectangular
aluminium cavity with a TE101 mode resonance frequency of 9.053 GHz. The
whole sample package is cooled in a dilution refrigerator to a base
temperature 10 mK. The transmon can be considered as an artificial atom
located in a cavity and the dynamics of the system is generally described by
the theory of circuit QED \cite{Blais,wallraff_qed,Koch,You_qed}. We
designed the energy level of the transmon qubit to let the system work in
the dispersive region \cite{You_qed}. The quantum states of the transmon
qubit can be controlled by microwaves. IQ mixers combined with 1 GHz
arbitrary wave generator (AWG) are used to modulate the amplitude,
frequency, and phase of microwave pulses. To readout qubit states, we used
an ordinary microwave heterodyne setup. The output microwave is
pre-amplified by a HEMT at the 4 K stage in the dilution refrigerator and
further amplified by two low-noise amplifiers at room temperature. The
microwave is then heterodyned into 50 MHz and collected by ADCs. The readout
is performed with a \textquotedblleft high power readout" scheme \cite%
{Reed_readout}. We sent in a strong microwave on-resonance with the cavity,
the transmitted amplitude of the microwave reflects the state of the
transmon due to the non-linearity of the cavity QED system.

According to circuit QED theory, the coupled transmon qubit and cavity
exhibit anharmonic multiple energy levels. In our experiments, we used the
lowest three energy levels, $|0\rangle $, $|1\rangle ,$ and $|2\rangle $, as
shown in Fig.~\ref{demon}c. The two states $|1\rangle $ and $|2\rangle $
behave as an artificial spin-1/2 particle, whose three components may be
denoted by the three Pauli matrices $\sigma_{1,2,3}$ which can couple with
the microwave fields. $|0\rangle $ is chosen as an ancillary level to probe
the energy spectrum of the simulated system. The transition frequencies
between different energy levels are $\omega _{10}/2\pi =$ 7.172 GHz, $\omega
_{21}/2\pi$ = 6.831 GHz, respectively, which are independently determined by
saturation spectroscopies. The energy relaxation time of the qubit is $%
T_1\sim$7 $\mu s$, the dephasing time is $T^*_2\sim$6 $\mu s$. When we apply
microwave drives along the $x$, $y$, and $z$ directions, the effective
Hamiltonian of the qubit in the rotating frame may be written as ($\hbar =1$
for simplicity)

\begin{equation}
\hat{H}=\sum_{i=1}^{3}\Omega _{i}\sigma _{i}/2,
\end{equation}%
where $\Omega _{1}$ $(\Omega _{2})$ corresponds to the frequency of Rabi
oscillations along the x (y) axis on the Bloch sphere, which is continuously
adjustable by changing the amplitude and phase of the microwave applied to
the system. $\Omega _{3}=\omega _{21}-\omega_m ,$ is determined by the
detuning between the system energy level spacing $\omega _{21}$ and
microwave frequency $\omega_m $. By carefully designing the waveform of the
AWG, we can modulate the frequency, amplitude, and phase of the microwave.
In our experiment, we first calibrated the parameters $\Omega _{1},$ $\Omega
_{2},$ and $\Omega _{3}$ using Rabi oscillations and Ramsey fringes, and
then designed the microwave amplitude, frequency and phase to set $\Omega
_{1}=\Omega (\sin k_y \cos k_z - \sin k_x \sin k_z),$ $\Omega _{2}(k_{x})=0,$
$\Omega _{3}(k_{y})=\Omega (2 \cos k_x + 2 \cos k_y +\chi)$, point-by-point
in the parameter space, with $\Omega =$ 10 MHz being the energy unit here.
By mapping the parameter space of the driving two-level system to the $%
\mathbf{k}$-space of a lattice Hamiltonian system, we have realized Eq.~%
\eqref{Lattice-model} exactly.

\begin{figure}[tbph]
\includegraphics[width=7.5cm]{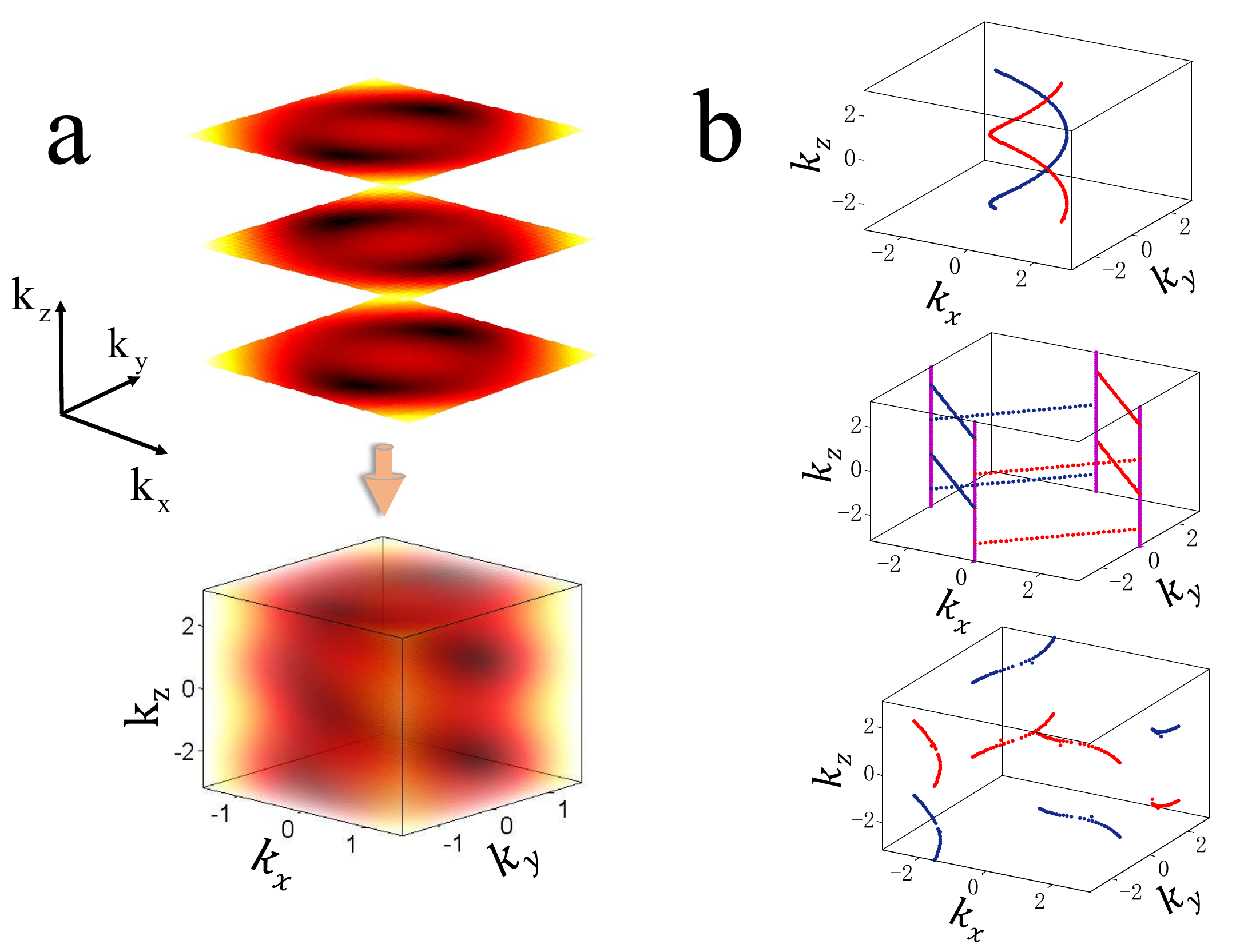}
\caption{(Color online) \textbf{a.} Measurement of the band structure of the
Hopf-link semimetal in the first Brillouin zone. Top panel: Contour plots of
the energy gap with varying $k_{z}$ gradually in the range of $[-\protect\pi %
,\protect\pi ].$ Bottom panel: By collecting all these contour plots
together, we obtain the nodal lines of the Hopf-link semimetal. To image the
gapless band structure clearly, we set the range of ($k_x,k_y$) as $[-%
\protect\pi/2 ,\protect\pi/2 )\times [-\protect\pi/2,\protect\pi/2)$.
\textbf{b.} Nodal lines obtained from the measured energy spectrum for
various $\protect\chi$. From top to bottom: $\protect\chi$ = -3, 0, and 2,
respectively.}
\label{nodal-Lines}
\end{figure}

It was predicted \cite{chen_hopf} that $\chi $ plays a crucial role in the
realization of the double-helix nodal lines in topological semimetal. We
first examined the transition of the Hopf-link band structure with the
change of $\chi $. Starting from $\chi =-3$, we measured the entire energy
spectrum of the system over the first Brillouin zone (BZ). The energy
spectrum is measured with a spin injection technique \cite{Tan_pt}. The
system is always initialized in $|0\rangle $. For a preset ($k_x,
k_y,k_z)\in [-\pi ,\pi )\times [-\pi,\pi)\times [-\pi,\pi)$, the construct
microwave pulse drives the two-level system to obtain the Hamiltonian in
Eq.~(1). A probe microwave pulse is then sent in. When the frequency of the
probe microwave matches the energy spacing between the eigenenergy of the
Hamiltonian and $|0\rangle $, the system will be excited to the eigenstate.
A resonant peak of microwave absorption can then be observed. The frequency
of the resonant peak represent the eigenenergy. We then gradually changed $%
k_{x}$, $k_{y}$, and $k_{z}$, collecting all the eigenenergies from the
resonant peaks of the spectrum. We can extract the band structure of the
semimetal in the first BZ. The zero energy points of the band structure can
be directly imaged as nodal points. A chain of connected nodal points forms
a nodal line. In Fig.~\ref{nodal-Lines}b, we plot the nodal lines in the
first BZ. A feature of the Hopf-link topological semimetal, which is a
double-helix structure, is clearly observed, indicating that we have
successfully realized the Hopf-link topological semimetal. In addition, the
positions of the nodal lines agree well with the theoretical calculation of
Eq.~\eqref{Lattice-model} with $\lambda=-3$. With increasing $\chi$, the
surface $S_2$ expands, leading to a shape change of double-helix. At $\chi$
= 0, $S_2$ touches the boundary of the Brillouin zone, and the double-helix
deforms to a cubic-like shape (middle panel in Fig. \ref{nodal-Lines}b). At $%
\chi$ = 2, $S_2$ opens, forming a cylinder centered at ($\pi,\pi$). The
intersection of $S_x$ and $S_y$ remains a double-helix shape (as shown in
the bottom panel of Fig.~\ref{nodal-Lines}b).

\begin{figure}[tbph]
\includegraphics[width=7.5cm]{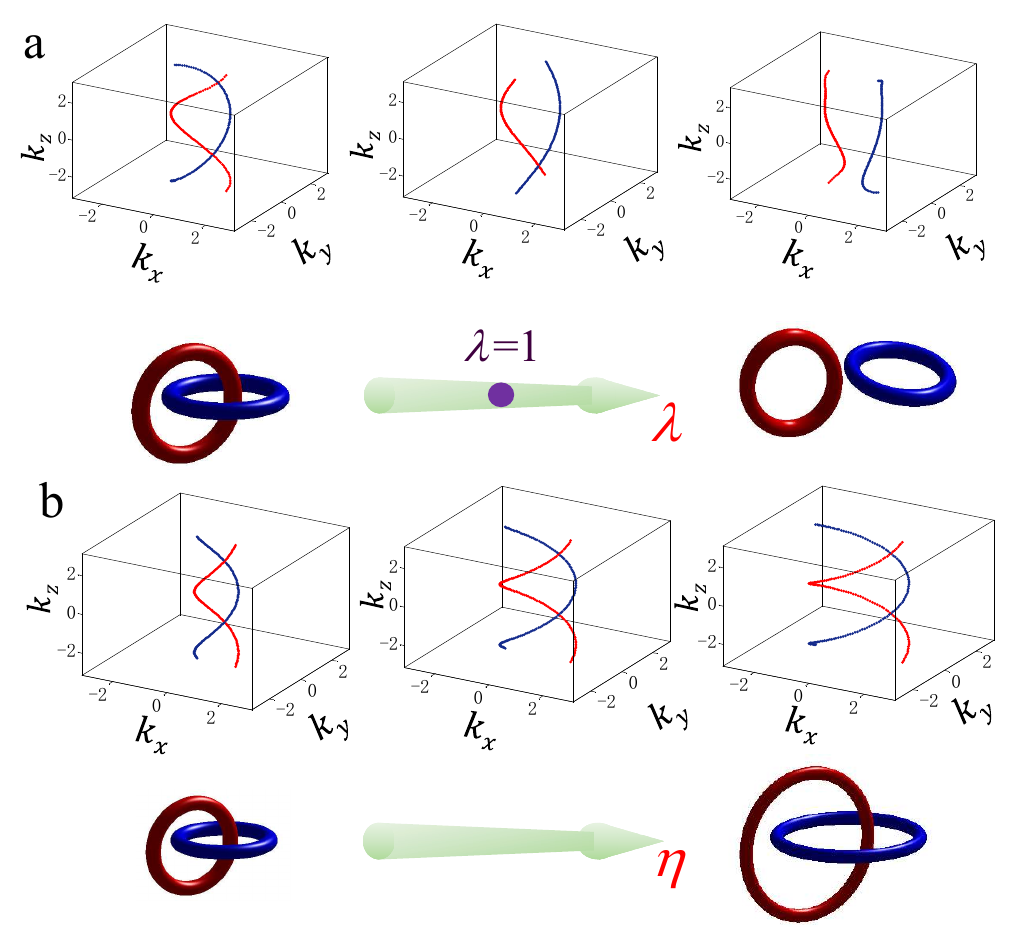}
\caption{(Color online) \textbf{a}, Link-unlink transition of the Hopf-link
semimetal after we added $H_1^{\prime}$ = $\protect\lambda \sin k_y \protect%
\sigma_{1}$ to the Hamiltonian in Eq.~(\protect\ref{Lattice-model}). From
left to right: nodal lines with $\protect\lambda$ = 0.5, 1, and 1.5,
respectively. As predicted, $\protect\lambda$ = 1 is the critical point,
where two nodal rings only touch at one point. \textbf{b}, Stability of the
Hopf-link against the perturbation of $H_{2}^{\prime}= \protect\eta \protect%
\sigma _{3}$. From left to right: nodal lines with $\protect\eta$ = -0.5,
0.5, and 1.5, respectively. }
\label{transition}
\end{figure}

Furthermore, one can take advantage of the full tunability of the
superconducting circuit to demonstrate the topological transition and
stability. First, we introduce an additional term $H_1^{\prime}$ = $\lambda
\sin k_y \sigma_{1}$ to manipulate the topological link-unlink transition
\cite{chen_hopf}. Here $H_1^{\prime}$ is in units of $\Omega$, and $\lambda$
is an adjustable parameter. Shown in Fig. \ref{transition}a are nodal-line
structures for different $\lambda$. When $\lambda <$ 1, two nodal lines are
linked. As we increased $\lambda $ to larger than 1, the two nodal lines
become separated, forming two topologically unlinked isolation loops. For $%
\lambda$ = 1, the projection of the nodal lines connect at a single point.
We can consider this point as the critical point of the transition.

Since the Hamiltonian in Eq.~(1) commutes with $PT$, the Hopf-link is
protected by the joint space-time symmetry. In order to verify this, we add
a term $H_{2}^{\prime}= \eta \sigma _{3}$ (with $\eta$ in units of $\Omega$)
to the Hamiltonian in Eq.~(\ref{Lattice-model}). The modified Hamiltonian
violates single $P$ and $T$ symmetry but preserves the joint $PT$ symmetry.
It is found that with increasing $\eta$, the band structure is distorted
dramatically, and the positions as well as neighborhood geometries of the
band-crossing lines are changed significantly. Nevertheless, these nodal
lines are persistently present in the BZ without opening any gap, thus
forming a double-helix, as shown in Fig.~\ref{transition}b. This strongly
supports that the Hopf-link structure is protected by the $PT$ symmetry.

\begin{figure}[tbph]
\includegraphics[width=7.5cm]{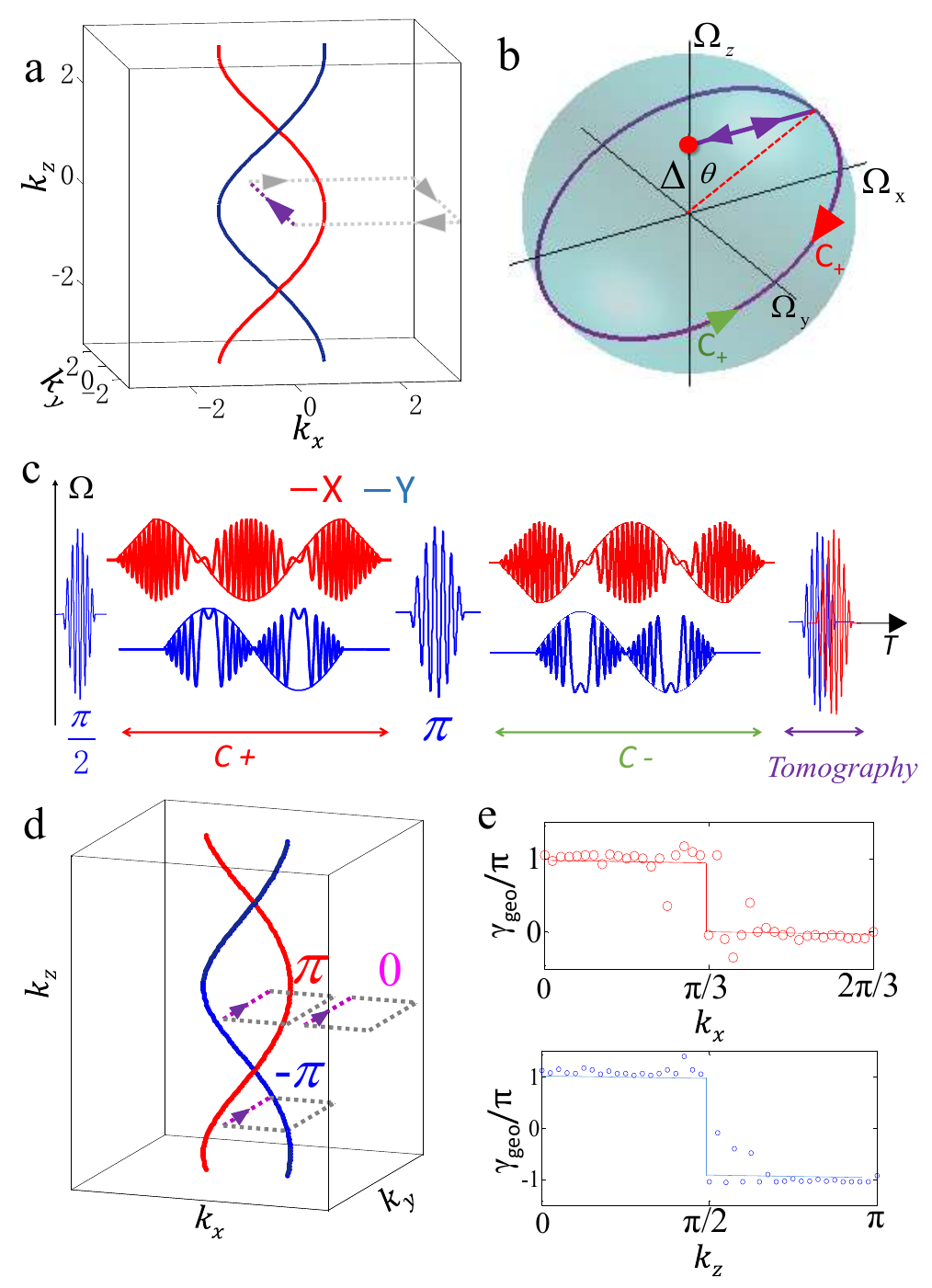}
\caption{(Color online) \textbf{a}, Schematic of an example of closed path
(dashed line) in the first Brillouin zone to accumulate Berry phase, from
which the linking number can be characterized. \textbf{b}, Evolving path in
parameter space of qubit mapped from momentum space in $\mathbf{a}$. \textbf{%
c}, Schematic of time profile to probe Berry phase accumulated from the
evolution in \textbf{b}. \textbf{d}, Dependance of the Berry phase carried
on the loop of the closed path. There are three typical values: $\protect\pi$%
, $-\protect\pi$, and $0$. \textbf{e}, Berry phase measured as a function of
$k_x$ (top panel) and $k_z$ (down panel), respectively. }
\label{berry-phase}
\end{figure}

Besides the band structure, the topological invariant quantity is another
feature associated with topological phenomena. To characterize the linking
number of the Hopf-link, which describes the connection of the nodal lines,
we detect the Berry phase carried by a closed path enclosing or disclosing
nodal lines in the parameter space from an adiabatic procedure \cite%
{abanin_zak,xiao_berry}. For the dashed loop showed in Fig.~\ref{berry-phase}%
a, we design a closed path enclosing the Hopf-link chains in Brillouin zone
to probe the Berry phase. The entire adiabatic evolution consists of two
parts. One path, marked by the purple dashed line, which threads the link
loop, contributes a Berry phase $\pi $. The other part, marked with grey
dashed lines, contributes a null Berry phase. However, the loop in Fig. \ref%
{berry-phase}a is in momentum space. We have to map it from momentum space
to the parameter space of the superconducting qubit, as illustrated in Fig.~\ref{berry-phase}b.
Theoretically, the evolution path denoted by the purple
dashed line is topologically equivalent to a circle along the meridian of
the Bloch sphere. Without loss of generality, we design a geodesic to
replace the original path by choosing $\{\Omega _{x},\Omega _{y},\Omega
_{z}\}=\{\sin \theta \cos \phi +\Lambda _{1},\Lambda _{2}\sin \phi ,\cos
\theta \cos \phi +\Lambda _{1}\}$, where $\theta \in \lbrack 0,\pi ]$ and $%
\phi \in \lbrack 0,2\pi ]$ are spherical coordinates. By tuning the
parameters $\Lambda _{1}$ and $\Lambda _{2}$, we can implement any closed
path in the Brillouin zone. To detect the accumulated Berry phase after time
evolution, we used the Ramsey fringe interference technique \cite%
{Leek_berry,Tan_geo}. This adiabatic approach has been demonstrated to be a
convenient method to measure the Berry phase in superconducting circuits. If
one prepares the qubit in a superposition state $(|0\rangle +|1\rangle )/%
\sqrt{2}$, the evolution of $|0\rangle $ and $|1\rangle $ will acquire a
relative Berry phase $\gamma _{\mathrm{geo}}$, which equals to 1/2 of the
solid-angle enclosed by the circle denoted as $C_{\pm }$. The sign depends
on the path direction. The relative phase of the quantum state after the
evolution is extracted as $\Phi _{\mathrm{total}}$ = $\arctan (\sigma
_{y}/\sigma _{x})$, which equals four times of the Berry phase. This agrees
with the theoretical prediction, confirming the topological properties of
the nodal rings. As shown in Fig.~\ref{berry-phase}d, by adjusting the
parameters $\Lambda _{1}$ and $\Lambda _{2}$, we can move the position of
the closed path in \textbf{k}-space. By moving the purple dashed line along
the $k_{x}$ axis on the $k_{z}=0$ plane, the originally designed closed loop
changes from enclosing to disclosing the red nodal line. The geometric phase
measured then switches from $\pi $ to 0 abruptly at $k_{x}=\pi /3$,
indicating that the closed path in the qubit parameter space is no longer a
geodesic circle at this critical point. Furthermore, by moving the loop
along $k_{z}$ on the $k_{x}=0$ plane, the closed path changes from enclosing
a red nodal line to a blue nodal line. The Berry phase jumps from $\pi $ to $%
-\pi $ at $k_{z}=\pi /2$, corresponding to the Berry phase obtained along
the $C_{\pm }$ loop. Therefore, in the first Brillouin zone, the Berry phase
measured are well characterized topological connection of the nodal ring
\cite{chen_hopf}.

\section*{Discussion}

Experimentally realizing the Hopf-link band structures and thus
investigating related interesting topological properties in real condensed
matter systems are a challenge. Nodal lines with Hopf-chain structure have
not been observed in any multi-particle system so far. The lack of technique
for directly imaging the whole momentum-dependent electronic energy spectrum
also prohibits the fundamental research of the complex topological band
structure, noting only a part of the electronic spectra (or information of
Fermi surfaces/points) may be inferred from the angle-resolved photoemission
spectroscopy data (or quantum oscillation measurements) in bulk condensed
matter systems. Furthermore, it seems extremely difficult to tune the
parameters continuously for studying rich topological properties including
various topological quantum phase transitions in real materials. Therefore,
our simulations using artificial quantum systems, like superconducting
quantum circuits, provide faithful topological properties of the system,
which are useful to design related materials and devices.

\section*{Methods}

%

We used the Ramsey fringe interference technique to measure the Berry phase.
The schematic time profile of the measurement procedure is shown in Fig. \ref%
{berry-phase}c. At first, the qubit is initialized at $(|0\rangle +|1\rangle
)/\sqrt{2}$ by a $\pi _{y}/2$ pulse. Then, we ramp the Hamiltonian linearly
from $\{0,0,\Omega _{z}\}$ to $\{\Omega _{x},0,\Omega _{z}\}$, followed by a
traverse along the geodesic path $C_{+}$, by ramping the parameter $\phi
=2\pi t/T_{\mathrm{ramp}}$, where $T_{\mathrm{ramp}}$ = 400 ns. Since $%
\Omega $ is set as $2\pi \times $ 25 MHz, the adiabaticity in the evolution
is satisfied. Instantaneous spin-up and spin-down eigen-states of
Hamiltonian (denoted by $|\uparrow \rangle $ and $|\downarrow \rangle $ )
obtains the relative phases $\Phi _{c_{\pm }}=\pm \gamma _{\mathrm{geo}%
}+\phi _{\mathrm{dym}\uparrow (\downarrow )}$ respectively, where the Berry
phase $\gamma _{\mathrm{geo}}$ = $\pi $ in our experiments, and $\phi _{%
\mathrm{dym}\uparrow (\downarrow )}$ is the dynamical phase obtained in the
procedure. After a resonant spin-echo $\pi $ pulse is applied, the system
evolves along $C_{-}$, which has an opposite direction to $C_{+}$, acquiring
$\Phi _{c_{\mp }}=\mp \gamma _{\mathrm{geo}}+\phi _{\mathrm{dym}\downarrow
(\uparrow )}$. The states finally evolve to exp$[i(\phi _{\mathrm{dym}%
\uparrow }+\phi _{\mathrm{dym}\downarrow }+2\gamma _{\mathrm{geo}})]\Vert
\uparrow \rangle $ and exp$[i(\phi _{\mathrm{dym}\uparrow }+\phi _{\mathrm{%
dym}\downarrow }-2\gamma _{\mathrm{geo}})]\Vert \downarrow \rangle $,
therefore, the net relative phase gained during the complete procedure
equals $4\gamma _{\mathrm{geo}}$ \cite{Leek_berry}. At the end of the
evolution, we extract the phase of the qubit state by quantum state
tomography. To map different paths in the first Brillouin zone, $\Lambda _{1}
$ and $\Lambda _{2}$ are designed at various values. For instance, to move
the closed loop with fixed $k_{z}$ = 0 as shown in Fig. \ref{berry-phase}d, $%
\Lambda _{1}$ is varied as $(2\cos k_{x}-\lambda )/2$ and $\Lambda _{2}$ is
set as 1. 

\noindent \textbf{Acknowledgements:}

We would like to thank H.~Zhang for helpful suggestion, discussion, and
proof reading of the manuscript. We also thank F.~Nori for giving many helpful comments and suggestions after proof reading of the manuscript. This work was partly supported by the the
NKRDP of China (Grant No. 2016YFA0301802), NSFC (Grant No. 11274156, No.
11504165, No. 11474152, No. 61521001)


\begin{thebibliography}{99}
\bibitem{Hasan_RMP}  M.~Z. Hasan, C.~L. Kane, Colloquium: Topological
insulators.  \newblock {\it Rev. Mod. Phys.\/} {\bf 82}, 3045--3067 (2010).

\bibitem{Qi_RMP}  X.-L. Qi, S.-C. Zhang, Topological insulators and
superconductors.  \newblock {\it Rev. Mod. Phys.\/} {\bf 83}, 1057--1110
(2011).

\bibitem{XGWan-PRB11}  X.~Wan, A.~M. Turner, A.~Vishwanath, S.~Y. Savrasov,
Topological semimetal and  {F}ermi-arc surface states in the electronic
structure of pyrochlore  iridates.  \newblock {\it Phys. Rev. B\/} {\bf 83},
205101 (2011).

\bibitem{Konig_Hall}  M.~K\"{o}nig, S.~Wiedmann, C.~Brune, A.~Roth,
H.~Buhmann, L.~W. Molenkamp,  X.~Qi, S.~Zhang, Quantum spin {H}all insulator
state in $\mathrm{HgTe}$  quantum wells.  \newblock {\it Science\/} {\bf 318}%
, 766--770 (2007).

\bibitem{Kane_Z2}  C.~L. Kane, E.~J. Mele, ${Z}_{2}$ topological order and
the quantum spin {H}all  effect.  \newblock {\it Phys. Rev. Lett.\/} {\bf 95}%
, 146802 (2005).

\bibitem{Lu-Science15}  L.~Lu, Z.~Wang, D.~Ye, L.~Ran, L.~Fu, J.~D.
Joannopoulos, M.~Solja{\v c}i{\'c},  Experimental observation of {W}eyl
points.  \newblock {\it Science\/} {\bf 349}, 622--624 (2015).

\bibitem{Xu-Science15}  S.-Y. Xu, I.~Belopolski, N.~Alidoust, M.~Neupane,
G.~Bian, C.~Zhang, R.~Sankar,  G.~Chang, Z.~Yuan, C.-C. Lee, S.-M. Huang,
H.~Zheng, J.~Ma, D.~S. Sanchez,  B.~Wang, A.~Bansil, F.~Chou, P.~P.
Shibayev, H.~Lin, S.~Jia, M.~Z. Hasan,  Discovery of a {W}eyl fermion
semimetal and topological {F}ermi arcs.  \newblock {\it Science\/} {\bf 349}%
, 613--617 (2015).

\bibitem{LiuDirac}  Z.~Liu, J.~Jiang, B.~Zhou, Z.~Wang, Y.~Zhang, H.~Weng,
D.~Prabhakaran, S.~Mo,  H.~Peng, P.~Dudin, A stable three-dimensional
topological  {D}irac semimetal $\mathrm{Cd}_3\mathrm{As}_2$.  \newblock {\it
Nature materials\/} {\bf 13}, 677--681 (2014).

\bibitem{LvWeyl}  B.~Q. Lv, H.~M. Weng, B.~B. Fu, X.~P. Wang, H.~Miao,
J.~Ma, P.~Richard, X.~C.  Huang, L.~X. Zhao, G.~F. Chen, Z.~Fang, X.~Dai,
T.~Qian, H.~Ding,  Experimental discovery of {W}eyl semimetal $\mathrm{TaAs}$%
.  \newblock {\it Phys. Rev. X\/} {\bf 5}, 031013 (2015).

\bibitem{bian_nodal}  G.~Bian, T.~Chang, R.~Sankar, S.~Xu, H.~Zheng,
T.~Neupert, C.~Chiu, S.~Huang,  G.~Chang, I.~Belopolski, \textit{et~al.\/},
Topological nodal-line fermions in  spin-orbit metal $\mathrm{PbTaSe}_2$.  %
\newblock {\it Nat. Commun.\/} {\bf 7}, 10556--10556 (2016).

\bibitem{OskarWeyl}  O.~Vafek, A.~Vishwanath, Dirac fermions in solids: From
high-{T}c cuprates and  graphene to topological insulators and {W}eyl
semimetals.  \newblock {\it Annual Review of Condensed Matter Physics\/}
{\bf 5}, 83-112  (2014).

\bibitem{YuDiracline}  R.~Yu, H.~Weng, Z.~Fang, X.~Dai, X.~Hu, Topological
node-line semimetal and  dirac semimetal state in antiperovskite $\mathrm{Cu}%
_{3}\mathrm{PdN}$.  \newblock {\it Phys. Rev. Lett.\/} {\bf 115}, 036807
(2015).

\bibitem{KimDiracline}  Y.~Kim, B.~J. Wieder, C.~L. Kane, A.~M. Rappe, Dirac
line nodes in  inversion-symmetric crystals.  \newblock {\it Phys. Rev.
Lett.\/} {\bf 115}, 036806 (2015).

\bibitem{burkov_nodal}  A.~Burkov, M.~D. Hook, L.~Balents, Topological nodal
semimetals.  \newblock {\it Physical Review B\/} {\bf 84} (2011).

\bibitem{moore_hopf}  J.~E. Moore, Y.~Ran, X.~Wen, Topological surface
states in three-dimensional  magnetic insulators.  \newblock {\it Phys. Rev.
Lett.\/} {\bf 101}, 186805 (2008).


\bibitem{Neupane-NC15}  M.~Neupane, S.-Y. Xu, R.~Sankar, N.~Alidoust,
G.~Bian, C.~Liu, I.~Belopolski,  T.-R. Chang, H.-T. Jeng, H.~Lin, A.~Bansil,
F.~Chou, M.~Z. Hasan, Observation  of a three-dimensional topological {D}%
irac semimetal phase in high-mobility  $\mathrm{Cd}_3\mathrm{As}_2$.  %
\newblock {\it Nat. Commun.\/} {\bf 5}, 3786 (2014).

\bibitem{chen_hopf}  W.~Chen, H.-Z. Lu, J.-M. Hou, Topological semimetals
with a double-helix nodal  link.  \newblock {\it Phys. Rev. B\/} {\bf 96},
041102 (2017).

\bibitem{berry_berry}  M.~V. Berry, Quantal phase factors accompanying
adiabatic changes.  \newblock {\it Proceedings of The Royal Society of
London\/} {\bf 392}, 45--57  (1984).

\bibitem{chang_weyl-link}  P.-Y. Chang, C.-H. Yee, Weyl-link semimetals.  %
\newblock {\it Phys. Rev. B\/} {\bf 96}, 081114 (2017).

\bibitem{yan_nodal-link}  Z.~Yan, R.~Bi, H.~Shen, L.~Lu, S.-C. Zhang,
Z.~Wang, Nodal-link semimetals.  \newblock {\it Phys. Rev. B\/} {\bf 96},
041103 (2017).

\bibitem{Tan_pt}  X.~Tan, Y.~Zhao, Q.~Liu, G.~Xue, H.~Yu, Z.~D. Wang, Y.~Yu,
Realizing and  manipulating space-time inversion symmetric topological
semimetal bands with  superconducting quantum circuits.  \newblock {\it npj
Quantum Materials\/} {\bf 2}, 60 (2017).

\bibitem{paik_3d}  H.~Paik, D.~I. Schuster, L.~S. Bishop, G.~Kirchmair,
G.~Catelani, A.~P. Sears,  B.~R. Johnson, M.~J. Reagor, L.~Frunzio, L.~I.
Glazman, S.~M. Girvin, M.~H.  Devoret, R.~J. Schoelkopf, Observation of high
coherence in {J}osephson  junction qubits measured in a three-dimensional
circuit {QED} architecture.  \newblock {\it Phys. Rev. Lett.\/} {\bf 107},
240501 (2011).

\bibitem{devoret_3d}  I. Buluta, F. Nori, Quantum Simulators.  \newblock
{\it Science\/} {\bf 326}, 108--111 (2009).

\bibitem{shankar_3d}  K. W. Murch, S. J. Weber, C. Macklin, and I. Siddiqi.
Observing single quantum trajectories of a superconducting quantum bit.  %
\newblock {\it Nature\/} \textbf{502}, 211 (2013).

\bibitem{wang_3d}  Z. Kim, B. Suri, V. Zaretskey, S. Novikov, K. D. Osborn.
A. Mizel, F. C. Wellstood, and B. S. Palmer, Decoupling a Cooper-pair box
to enhance the lifetime to 0.2 ms.  \newblock {\it Phys. Rev. Lett.\/}
\textbf{106}, 120501  (2011).

\bibitem{Blais}  X. Gu, A.F. Kockum, A. Miranowicz, Y.X. Liu, F. Nori,
Microwave photonics with superconducting quantum circuits.  \newblock {\it
Physics Reports\/} {\bf 718-719}, 1-102 (2017).

\bibitem{You_qed}  J.~Q. You, F.~Nori, Quantum information processing with
superconducting qubits  in a microwave field.  \newblock {\it Phys. Rev.
B\/} {\bf 68}, 064509 (2003).

\bibitem{wallraff_qed}  A.~Wallraff, D.~I. Schuster, A.~Blais, L.~Frunzio,
R.~Huang, J.~Majer,  S.~Kumar, S.~M. Girvin, R.~J. Schoelkopf, Strong
coupling of a single photon  to a superconducting qubit using circuit
quantum electrodynamics.  \newblock {\it Nature\/} {\bf 431}, 162 (2004).

\bibitem{Koch}  J.~Q. You, F. Nori,  Atomic physics and quantum optics using
superconducting circuits.  \newblock {\it Nature\/} {\bf 474}, 589 (2011).

\bibitem{Reed_readout}  M.~D. Reed, L.~DiCarlo, B.~R. Johnson, L.~Sun, D.~I.
Schuster, L.~Frunzio,  R.~J. Schoelkopf, High-fidelity readout in circuit
quantum electrodynamics  using the {J}aynes-{C}ummings nonlinearity.  %
\newblock {\it Phys. Rev. Lett.\/} {\bf 105}, 173601 (2010).

\bibitem{abanin_zak}  D.~A. Abanin, T.~Kitagawa, I.~Bloch, E.~Demler,
Interferometric approach to  measuring band topology in 2{D} optical
lattices.  \newblock {\it Phys. Rev. Lett.\/} {\bf 110}, 165304 (2013).

\bibitem{xiao_berry}  D.~Xiao, M.~Chang, Q.~Niu, Berry phase effects on
electronic properties.  \newblock {\it Rev. Mod. Phys.\/} {\bf 82},
1959--2007 (2010).

\bibitem{Leek_berry}  P.~J. Leek, J.~M. Fink, A.~Blais, R.~Bianchetti, M.~G{%
\"o}ppl, J.~M. Gambetta,  D.~I. Schuster, L.~Frunzio, R.~J. Schoelkopf,
A.~Wallraff, Observation of  {B}erry{\textquoteright}s phase in a
solid-state qubit.  \newblock {\it Science\/} {\bf 318}, 1889--1892 (2007).

\bibitem{Tan_geo}  X.~Tan, D.-W. Zhang, Z.~Zhang, Y.~Yu, S.~Han, S.-L. Zhu,
Demonstration of  geometric {L}andau-{Z}ener interferometry in a
superconducting qubit.  \newblock {\it Phys. Rev. Lett.\/} {\bf 112}, 027001
(2014).
\end{thebibliography}
\end{document}